\documentclass[conference]{IEEEtran}
\IEEEoverridecommandlockouts
\usepackage{cite}
\usepackage{amsmath,amssymb,amsfonts}
\usepackage{algorithmic}
\usepackage{graphicx}
\usepackage{textcomp}
\usepackage{xcolor}
\usepackage{subfigure}
\usepackage{url}
\usepackage{comment}
\def\BibTeX{{\rm B\kern-.05em{\sc i\kern-.025em b}\kern-.08em
    T\kern-.1667em\lower.7ex\hbox{E}\kern-.125emX}}
\begin{document}

\title{Pricing Mechanisms versus Non-Pricing Mechanisms for Demand Side Management in Microgrids\\

\thanks{This research is partly supported by (1) Academy of Finland via: (a) EnergyNet Fellowship n.321265/n.328869/n.352654, and (b) X-SDEN project n.349965/349966; and (2) EU MSCA Project "COALESCE" under Grant Number 101130739.}
}

\author{\IEEEauthorblockN{1\textsuperscript{st} Cássia Nunes Almeida}
\IEEEauthorblockA{\textit{School of Energy Systems} \\
\textit{Lappeenranta--Lahti University of Technology}\\
Lappeenranta, Finland \\
cassia.almeida@lut.fi}
\and
\IEEEauthorblockN{2\textsuperscript{nd} Arun Narayanan}
\IEEEauthorblockA{\textit{School of Energy Systems} \\
\textit{Lappeenranta--Lahti University of Technology}\\
Lappeenranta, Finland \\
arun.narayanan@lut.fi}
\and
\IEEEauthorblockN{3\textsuperscript{rd} Majid Hussain}
\IEEEauthorblockA{\textit{School of Energy Systems} \\
\textit{Lappeenranta--Lahti University of Technology}\\
Lappeenranta, Finland \\
majid.hussain@lut.fi}
\and
\IEEEauthorblockN{4\textsuperscript{th} Pedro H. J. Nardelli}
\IEEEauthorblockA{\textit{School of Energy Systems} \\
\textit{Lappeenranta--Lahti University of Technology}\\
Lappeenranta, Finland \\
pedro.nardelli@lut.fi}
}

\maketitle

\begin{abstract}
In this paper, we compare pricing and non-pricing mechanisms for implementing demand-side management (DSM) mechanisms in a neighborhood in Helsinki, Finland. We compare load steering based on peak load-reduction using the profile steering method, and load steering based on market price signals, in terms of peak loads, losses, and device profiles. We found that there are significant differences between the two methods; the peak-load reduction control strategies contribute to reducing peak power and improving power flow stability, while strategies primarily based on prices result in higher peaks and increased grid losses. Our results highlight the need to potentially move away from market-price-based DSM to DSM incentivization and control strategies that are based on peak load reductions and other system requirements.
%This paper offers a comparative analysis of various DSM mechanisms applied across three distinct configuration settings within the context of the Profile Steering method. The study aims to evaluate the effectiveness of these strategies in reducing peak demand and optimizing energy consumption while minimizing the energy losses in the grid. The results and discussions are presented using a simulation tool called DEMKit, which demonstrates the application of DSM in a residential setting in a Helsinki neighborhood and the optimization of energy load at the transformer level. 

\end{abstract}


%%%%%%%%%%%%%%%%%%%%%%%%%%%%%%%%%%%%%%%%%%%%%%%%%%%%%%%%%%%%%%%%%%%%%%%%%%%%%%%%%%%%%%%%%%%
\section{Introduction}
%Finland has set an ambition climate target to reach carbon neutrality by 2035 and become carbon negative soon afterward. To reach this target, Finland has made tremendous progress by increasing shares of nuclear, hydro, and bio energy as well as onshore wind generation. In fact, Finland has one of the lowest levels of reliance on fossil fuels among the IEA member countries \cite{IEA2023}. In particular, Finland has aimed to modernize the traditional electricity system that has long provided a reliable and stable supply of electricity, but today, face important challenges such as environmental concerns, the need for carbon-neutral solutions, grid inefficiencies, and an ageing infrastructure. The Finnish electricity system now incorporates a greater share of renewable energy sources (RES), driven by the construction of wind farms, solar arrays, and renewable energy infrastructure.
Electricity systems nowadays incorporate a significant share of renewable energy sources (RES), driven by the construction of wind farms, solar arrays, and renewable energy infrastructure. However,  due to their intermittent nature and their decentralized integration, a higher share of RES introduces variability and unpredictability into the grid, leading to higher electricity supply fluctuations and peaks. As a result, new challenges have arisen, including a reduction in flexibility on the production side that intensifies the challenges of maintaining electricity balance. Production-side flexibility is typically achieved by energy storage, while demand-side flexibility is achieved by demand-side management (DSM) techniques \cite{FinGrid2022-2031}, leveraging the flexibility provided by electric vehicles, heat pumps, and smart home appliances whose electricity consumption and production can be shifted or adjusted. DSM is also strongly facilitated by the increasing adoption of advanced information and communications technologies, including smart meters. The application of DSM strategies contributes to the grid stability and helps to mitigate the risk of power outages and reductions in greenhouse gas emissions \cite{REIJNDERS202213281}.

A significant amount of literature has focused on pricing signals and mechanisms as a method to control appliances and achieve DSM, especially in countries with competitive electricity markets \cite{touma2021energy, panda2023comprehensive, elazab2024impacts}. These mechanisms use dynamic pricing strategies that include time-of-use tariffs, critical peak pricing, and real-time pricing \cite{panda2023comprehensive}. However, pricing signals have several disadvantages, often resulting in power quality problems and higher losses \cite{Gerards2015Demand,nardelli2018smart}. Instead, in many cases, it may be preferable to use reductions in peak loads explicitly as a control signal and incentivize customers to adjust their schedules accordingly based on the outcomes. Such a DSM that uses the desired (e.g., flat) power profiles as the objective could be especially useful to incentivize DSOs to immediately deploy green energy networks in the network, since the quality of electricity service is maintained. 

In this paper, we examine the impact of such a steering mechanism in Finnish conditions where the electricity infrastructure is ageing; the bulk of the electricity distribution network was originally constructed in the period 1950--70 and require urgent upgrades \cite{collan2024finnish}. We apply the Profile Steering method proposed in \cite{Gerards2015Demand} to a neighborhood in Finland, and compare the results with the usage of price as a signal. The essence of their Profile Steering method lies in the optimization of energy consumption profiles within households or buildings. This optimization is achieved through the utilization of advanced control algorithms, which take into account not only pricing dynamics but also network reliability and improved operational conditions. To provide a comprehensive comparison, we analyze the outcomes when steering was based on three different criteria: 100\% emphasis on pricing, 100\% prioritization of grid quality, and a balanced compromise between pricing and grid quality, with a 50\% weight assigned to each criteria. The results from the Profile Steering method are compared with a baseline scenario where all appliances operate without any control, initiating their operations as soon as they become available. This comparative analysis provides valuable insights into the impact of DSM mechanisms on grid utilization and energy efficiency within the Helsinki neighborhood. We also discuss how DSM strategies can be enhanced by the increasing availability of flexible loads. Together, these developments have created a more dynamic and responsive electricity grid, which is better equipped to handle the challenges of integrating RES.

\section{Methodology}\label{sec:PS}

Many DSM approaches use price steering signals to manage and optimize electricity demand. However, uniform pricing schemes can lead to load peaks being shifted in time instead of being reduced and control algorithms can even exacerbate the problem. Differentiated dynamic pricing, which uses different price signals for individual houses, can alleviate the problem but may still result in unbalanced loads and voltage problems. %Nonlinear pricing can also help to balance consumption.
The Profile Steering method, introduced by Gerards et al.\cite{Gerards2015Demand}, offers an innovative approach that considers future grid dynamics. This method strategically addresses the challenge of peak shaving at multiple hierarchical levels within the grid, offering increased flexibility while minimizing the necessity for substantial cable investments. Profile Steering operates through a decentralized control mechanism, wherein households receive desired power profiles as steering signals. These signals empower households to adapt and smooth their energy consumption patterns in accordance with the specified profiles. 

The Profile Steering algorithm is structured into two distinct phases: \textit{initialization} and \textit{iteration}. During the initialization phase, each household modifies its power profile to attain a balanced load distribution at the substation, which may involve achieving a flat profile or adhering to other load balancing schemes. When this phase is completed, the neighborhood exhibits a relatively uniform power profile. Then, in the iterative phase, the algorithm aggregates the power profiles of all households. These aggregated profiles are then analyzed to assess deviations from the ideal flat profile, quantified using the Euclidean distance metric. These deviations serve as steering signals for subsequent iterations, guiding further adjustments to the power profiles of individual households. This iterative process continues until the desired load balancing objective is successfully met. As a result, the Profile Steering method effectively mitigates peak loads, both locally and at the substations, ensures proper load balancing across the network and reduces energy losses within the grid. %As a result, the Profile Steering method effectively mitigates peak loads, both locally and at the substations, ensures proper load balancing across the network, maintains voltage levels within regulatory limits, and reduces energy losses within the grid.

This paper utilizes a software tool called DEMKit (Decentralized Energy Management Simulation and Demonstration Toolkit) to perform the simulations \cite{Hoogsteen2019DEMKit}. Developed at the University of Twente, DEMKit focuses on the decentralized management of multi-energy systems, including electricity, heat, hydrogen, using a model predictive control system. The control systems optimize local energy usage and reduce stress on energy distribution networks. DEMKit simulates microgrids in residential environments with a variety of assets, devices, and technologies, such as photovoltaics (PVs), batteries, electric vehicles (EVs), dishwashers, and washing machines. The architecture of DEMKit is based on a modular and cyber-physical systems design paradigm, with individual system components modeled and interacting with each other through predefined interfaces. This modular setup allows for the simulation of different control systems and device models, enabling evaluation of the effects of control systems on a physical grid. DEMKit is used in conjunction with ALPG (Artificial Load Profile Generator, \cite{Hoogsteen2016Generation}) to create power profiles of households and devices, providing explicit information on available flexibility. The ALPG is initialized with Gaussian-distributed parameters for households, individuals, and devices. Device presence is probabilistically determined; annual power consumption is calculated using truncated Gaussians, and household composition is established based on uniform distributions. An occupancy profile is also generated using a behavioral model. After setting up households and residents, devices like PV setups, battery storage, and BEVs are distributed according to predefined rules. %This ensures realistic device penetration and energy usage in households .Together, DEMKit and ALPG form a tool chain for smart grid studies of households and their devices.

%%%%%%%%%%%%%%%%%%%%%%%%%%%%%%%%%%%%%%%%%%%%%%%%%%%%%%%%%%%%%%%%%%%%%%%%%%%%%%%%%%%%%%%%%%%
\section{Results \& Discussions} \label{sec:Results}
Several simulations were conducted using DEMKit/ALPG to model a generic scenario of 10 houses in a street located in the Kumpula neighbourhood of Helsinki, Finland. The temperatures and global solar radiations data for Kumpula from 01/01/2017 to 31/12/2017 were obtained from the website of the Finnish Meteorological Institute \cite{FMI2023}. Finnish electricity prices from 2017 were used. A set of 10 houses were generated randomly, each equipped with PVs, controllable devices such as washing machines, dishwashers, ovens, electric vehicles, heat pumps, and batteries, and uncontrollable loads. The houses varied in terms of occupancy and included those with one person, couples, and families with single/dual parents and children.%, and dual-parent households with children. %The working status of each adult also varied and included unemployment, part-time employment, full-time employment, or retirement. 
The load profiles were generated using ALPG. %, and the simulation included additional electrification and local micro-generation using photovoltaic panels.
Some parameters for the model were configured as follows: the share of plug-in hybrid electric vehicles (PHEVs) and EVs were divided equally into 50\% each; photovoltaic panels (PV), battery, heat pump, combined heat and power also had 50\% penetration each. The other simulations parameters are presented in Table \ref{tab-Devices Parameters}. We chose the following steering configurations: 100\% control specifications, 100\% prices or to 50\% on control and 50\% on prices. The obtained results are compared to a base simulation without any steering. The objective is to minimize the risks of overloading the local grid by optimizing the energy load at the substation.

\begin{table}[htbp]
\caption{Devices Parameters}
\begin{center}
\begin{tabular}{|c|c|c|c|}
\hline
\textbf{Device} & \textbf{Parameter} & \textbf{Value} & \textbf{Unit} \\
\hline
EV & Capacity & 	42000	& Wh \\
\cline{2-4} 
  & Power & 		7400	& W \\
\hline
PHEV & Capacity & 	12000	& Wh \\
\cline{2-4}
 & Power & 	3700	& W \\
\hline
PV & Production & 	220		& kWh per year$^{\mathrm{a}}$ \\
\cline{2-4}
 & Efficiency & 		15 - 20		& \%$^{\mathrm{b}}$   \\
\hline
Battery & capacity & 	2000 - 12000 	& Wh   \\
\cline{2-4}
& power & 	3700 	& W   \\
\hline
\multicolumn{4}{l}{$^{\mathrm{a}}$average per m2 solar panel.}\\
\multicolumn{4}{l}{$^{\mathrm{b}}$of theoretical max.}
\end{tabular}
\label{tab-Devices Parameters}
\end{center}
\end{table}

%The system is set to run the profile steering method and the simulations with control are set to smart control, the use of fill method and event base method.

%The simulation starts in January 2017 and extends through December 2017, including all four seasons, allowing us to observe the dynamic energy consumption patterns that fluctuate over time. These variations in energy demand are driven by factors such as reduced solar incidence, leading to diminished photovoltaic generation. Conversely, when solar incidence rises, the PV generation also rises.

\begin{figure}
    \centering

    \subfigure[Power flow results without any steering conditions and when steering only on control strategies]{
        \includegraphics[width=0.47\textwidth]{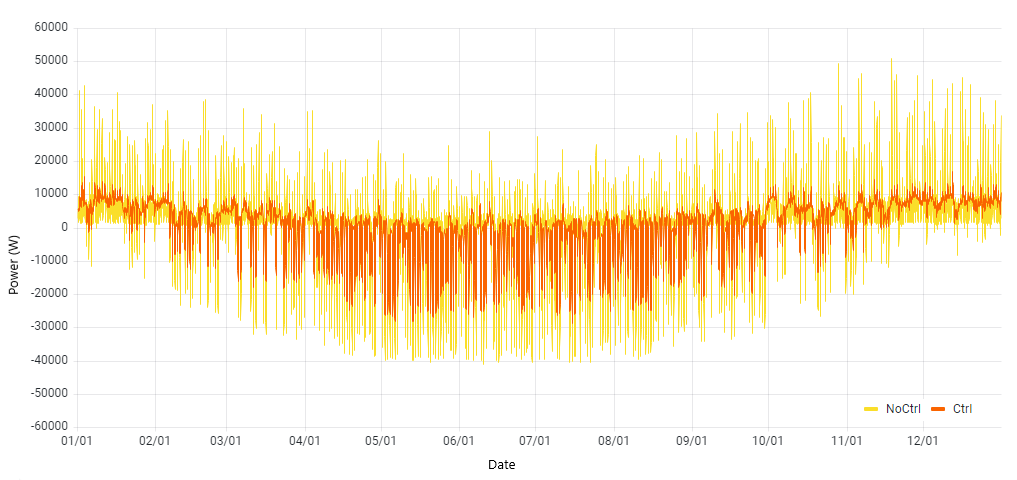} 
        \label{fig-street-10-houses-1.a}}
    \\
    \subfigure[Power flow results without any steering conditions and when steering only on prices conditions]{
        \includegraphics[width=0.47\textwidth]{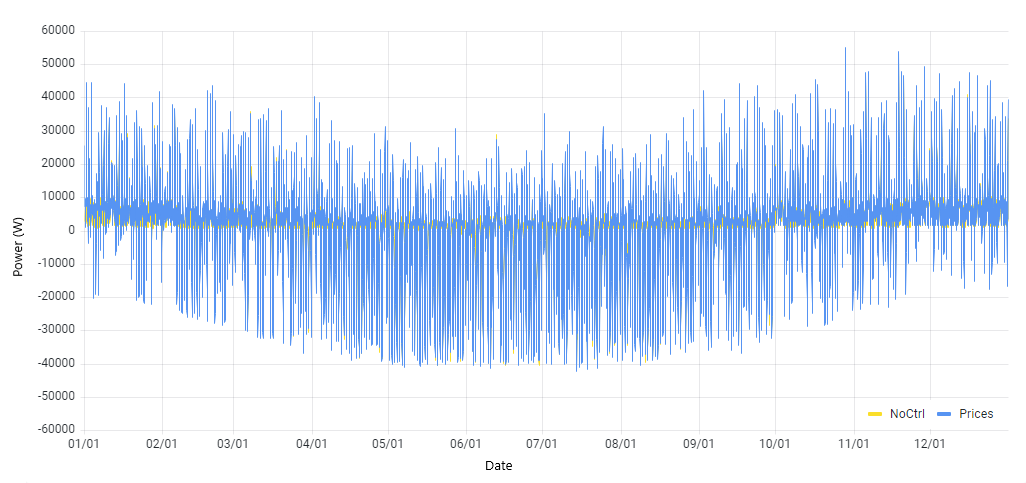} 
        \label{fig-street-10-houses-1.b}}
    \\
    \subfigure[Power flow results without any steering conditions and when steering on 50\% control strategies and 50\% prices]{
        \includegraphics[width=0.47\textwidth]{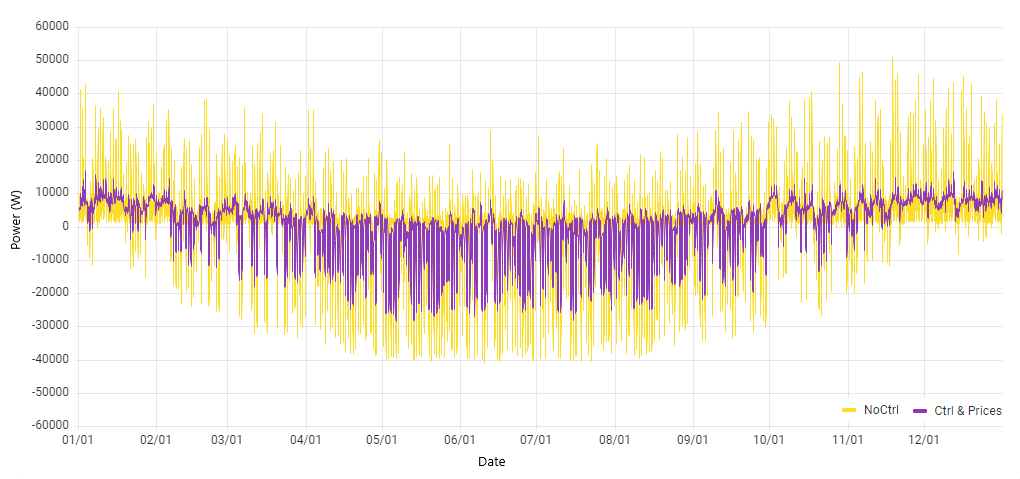} 
        \label{fig-street-10-houses-1.c}}
    \caption{Power flow results for simulations at the local lower voltage grid of the street with 10 houses at Helsink generated before and after being optimized at DEMKit \cite{Hoogsteen2019DEMKit} for the 2017 year}
    \label{fig-street-10-houses-1}
\end{figure}

\begin{figure}
    \centering
    \subfigure[Power Flow results in June without any steering, when steering only on Ctrl and when steering only on Prices and steering on Ctrl and Prices]{
        \includegraphics[width=0.47\textwidth]{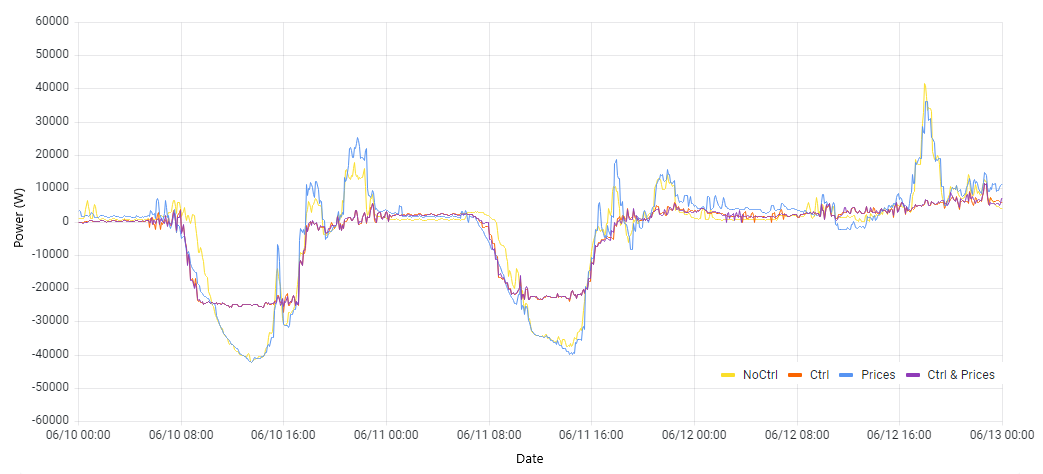} 
        \label{fig-street-10-houses-2J.a}}
    \\
    \subfigure[Losses results in June without any steering, when steering only on Ctrl, when steering only on Prices and when steering on Ctrl and Prices]{
        \includegraphics[width=0.47\textwidth]{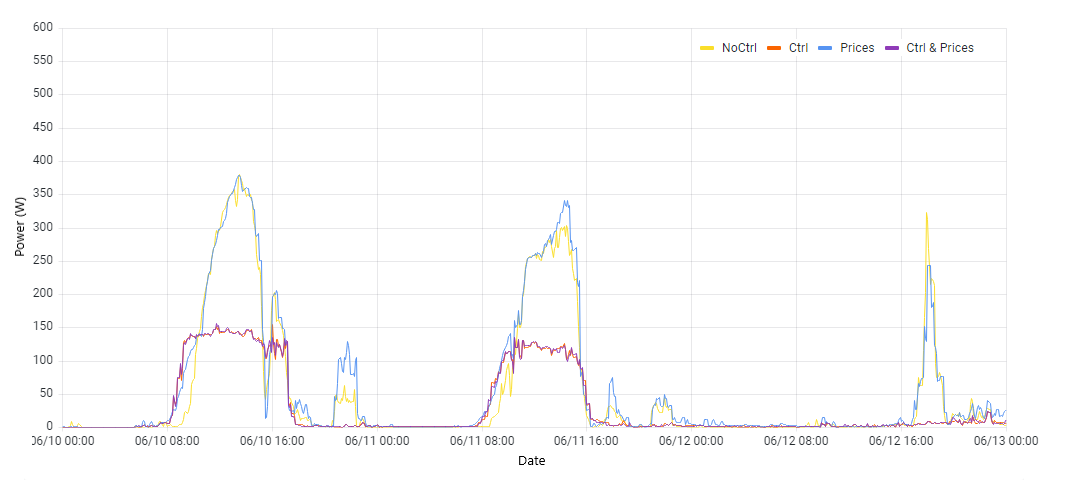} 
        \label{fig-street-10-houses-2J.b}}
    \caption{Power profiles and Losses results for the load flow at the local lower voltage grid of the street with 10 houses at Helsinki generated before and after being optimized at DEMKit \cite{Hoogsteen2019DEMKit} for 3 days in June of 2017}
    \label{fig-street-10-houses-2J}
\end{figure}

\begin{figure}
    \centering
    \subfigure[Power Flow results in December without any steering, when steering only on Ctrl and when steering only on Prices and steering on Ctrl and Prices]{
        \includegraphics[width=0.47\textwidth]{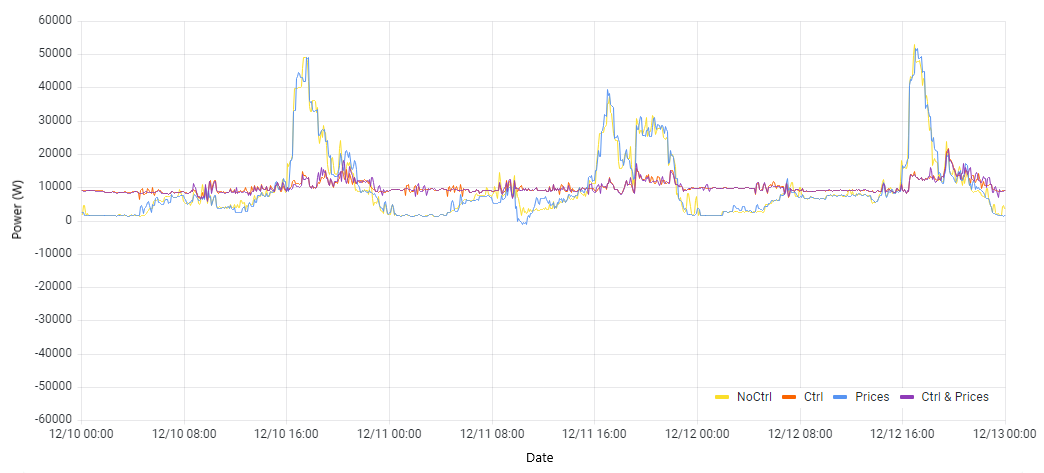} 
        \label{fig-street-10-houses-2D.a}}
    \\
    \subfigure[Losses results in December without any steering, when steering only on Ctrl, when steering only on Prices and when steering on Ctrl and Prices]{
        \includegraphics[width=0.47\textwidth]{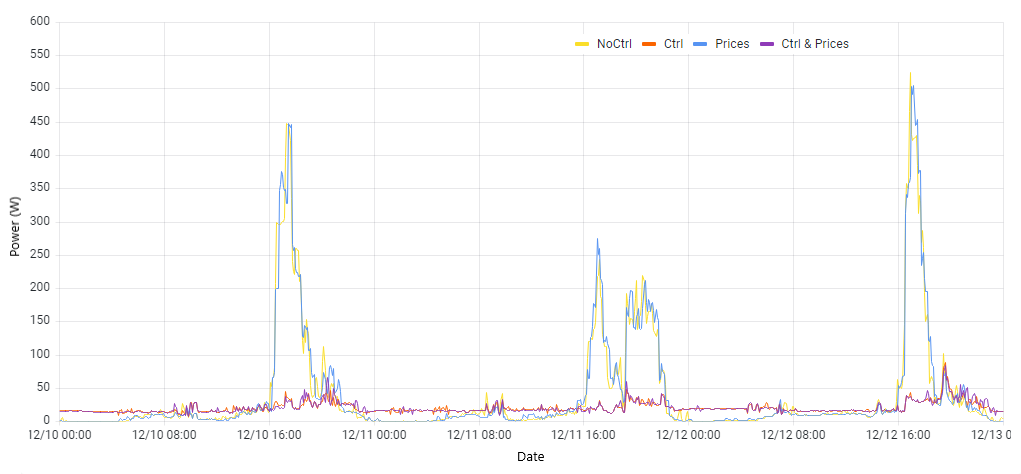} 
        \label{fig-street-10-houses-2D.b}}
    \caption{Power profiles and Losses results for the load flow at the local lower voltage grid of the street with 10 houses at Helsinki generated before and after being optimized at DEMKit \cite{Hoogsteen2019DEMKit} for 3 days in December of 2017}
    \label{fig-street-10-houses-2D}
\end{figure}

Figure \ref{fig-street-10-houses-1} illustrates the power flow within the local lower voltage grid of a 10-house in Helsinki neighborhood. The graph pictures the power distribution under different steering scenarios generated using the DEMKit tool \cite{Hoogsteen2019DEMKit} the the whole year of 2017. In Fig. \ref{fig-street-10-houses-1.a}, the results of the base simulation without any steering are compared to the outcomes of steering exclusively based on control strategies. The latter shows a reduction in peak values and a decrease in power flow variations compared to the base simulation. Fig. \ref{fig-street-10-houses-1.b} presents the power flow results under no steering conditions and when steering is exclusively based on pricing. Particularly, this scenario leads to higher peak values in the power flow compared to the base simulation. The grid operates under a more adverse conditions, resulting in increased losses, as observed in more detailed in Figures \ref{fig-street-10-houses-2J.b} and \ref{fig-street-10-houses-2D.b}. Fig. \ref{fig-street-10-houses-1.c} demonstrates power flow outcomes under no steering conditions and with a steering strategy comprising 50\% control strategies and 50\% price considerations. This simulation is closely similar to the results presented in Figure \ref{fig-street-10-houses-1.a}. we can see that the grid is more challenged in the hoter days, since they get more solar incidence and the photovoltaic generation increases.

Figure \ref{fig-street-10-houses-2J} provides insights into the power profiles (\ref{fig-street-10-houses-2J.a}) and losses (\ref{fig-street-10-houses-2J.b}) associated with the load flow within the local lower voltage grid of the 10-house street in Helsinki. The results for four different steering scenarios for a 3-day optimization in June 2017 are presented together, offering a more detailed perspective compared to the previous year-long simulation. The variations in power flow peaks are more pronounced for the steering only with prices. Further, the results when steering only on control strategies and steering on 50\% control strategies and 50\% prices are somewhat similar. In Figure \ref{fig-street-10-houses-2D}, the simulation results for December 2017 demonstrate patterns consistent with those observed in the June 2017 results presented in Figure \ref{fig-street-10-houses-2J}. Notably, the utilization of control strategies and a combination of 50\% control and 50\% price-based strategies results in lower power flow peaks and reduced losses compared to the June results under the same strategic steering. This detailed analysis highlights that the grid faces greater challenges and experiences increased losses on warmer days with higher solar incidence, consequently leading to higher photovoltaic generation than in the darker winter days.

\begin{figure}
    \centering
    \subfigure[Time-shiftable devices without any steering, when steering only on control strategies, when steering only on Prices and when steering on Ctrl and Prices]{
        \includegraphics[width=0.47\textwidth]{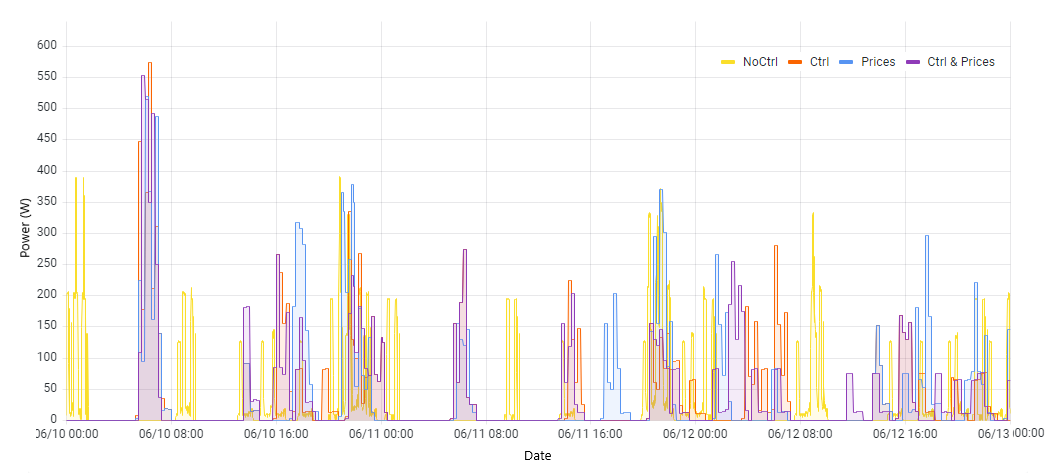} 
        \label{fig-street-10-houses-3.a}}
    \\
    \subfigure[Buffer-time-shiftable devices in the street of 10 houses without any steering, when steering only on control strategies, when steering only on Prices and when steering on Ctrl and Prices]{
        \includegraphics[width=0.47\textwidth]{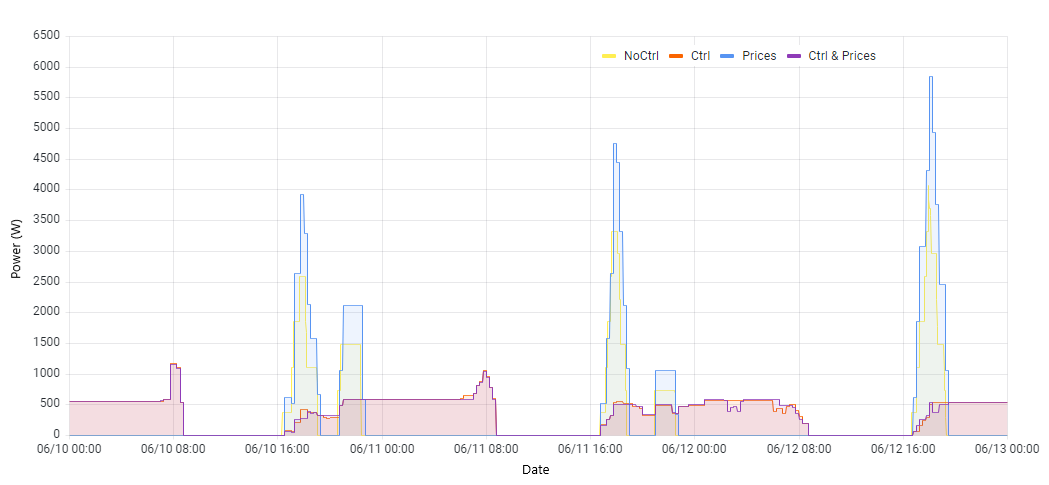} \label{fig-street-10-houses-3.b}}
    \caption{Power profiles results for two different set of devices from the street of 10 houses generated before and after being optimized at DEMKit \cite{Hoogsteen2019DEMKit} for 3 days in June of 2017}
    \label{fig-street-10-houses-3}
\end{figure}

Figure \ref{fig-street-10-houses-3} provides a comprehensive view of power profiles derived from two distinct device categories within a 10-house street in Helsinki. These profiles were generated before and after a 3-day optimization is applied using the DEMKit tool in June 2017. In Fig. \ref{fig-street-10-houses-3.a}, the aggregate power profile of time-shiftable devices in the street, including appliances like washing machines and dishwashers, is presented. Similarly, Fig. \ref{fig-street-10-houses-3.b} illustrates the cumulative power profile of buffer-time-shiftable devices, typified by electric vehicles with specific charging and discharging time frames (buffer devices). Within the time-shiftable devices graph, one can observe variations in equipment activation and deactivation times, with each strategic application yielding different operational periods. Notably, the majority of peaks are reduced in the strategies where steering primarily involves control strategies and a combination of 50\% control and 50\% pricing in relation to the results without any steering strategic, whereas the opposite occurs when steering primarily focuses on pricing strategies is applied. For buffer-time-shiftable devices, the most significant divergence is evident in the charging pattern curve of electric vehicles. This curve exhibits a more extended charging duration and a diminished peak when steering primarily incorporates control strategies and a combination of 50\% control and 50\% pricing, as opposed to the results without any steering strategy or price steering method that can have extremely high peak. These results also show that the charging profile of Electric Vehicles influences the power flow roughly ten times more than the sum of the other time-shiftable devices when the steering conditions are set for prices or without steering settings.

\begin{figure}
    \centering
    \subfigure[Devices of one house without any steering]{
        \includegraphics[width=0.47\textwidth]{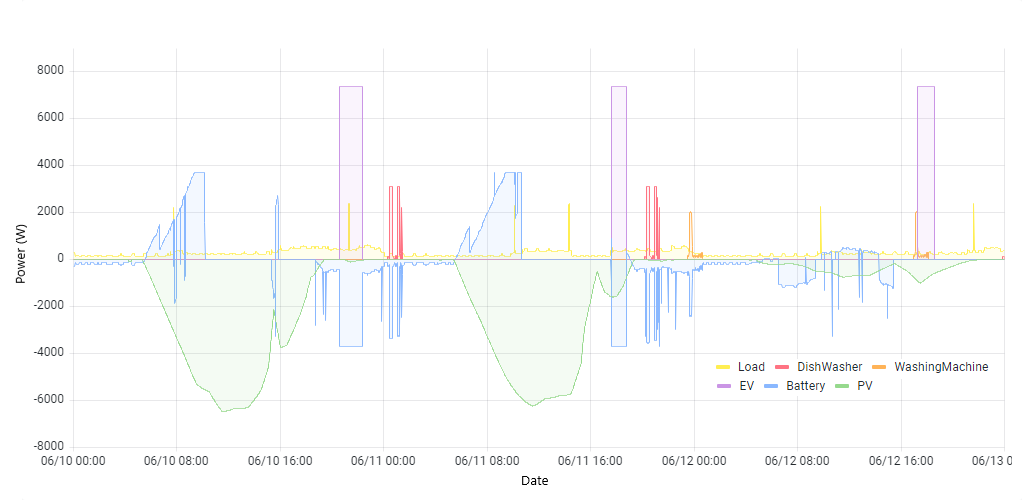} 
        \label{fig-street-10-houses-4.a}}
    \\
    \subfigure[Devices of one house when steering  only on control]{
        \includegraphics[width=0.47\textwidth]{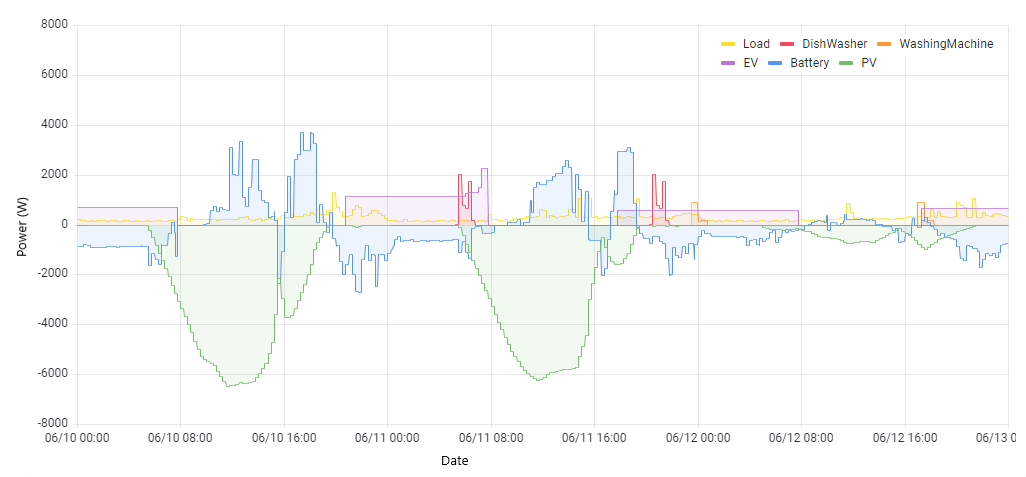} 
        \label{fig-street-10-houses-4.b}}
    \\
    \subfigure[Devices of one house when steering only on Prices]{
        \includegraphics[width=0.47\textwidth]{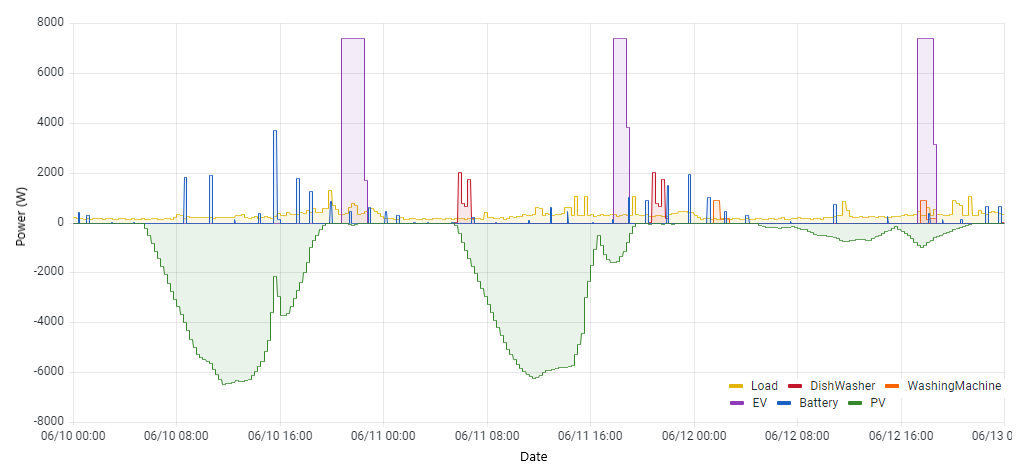} 
        \label{fig-street-10-houses-4.c}}
    \\
    \subfigure[Devices of one house when steering on 50\% Control and 50\% Price]{
        \includegraphics[width=0.47\textwidth]{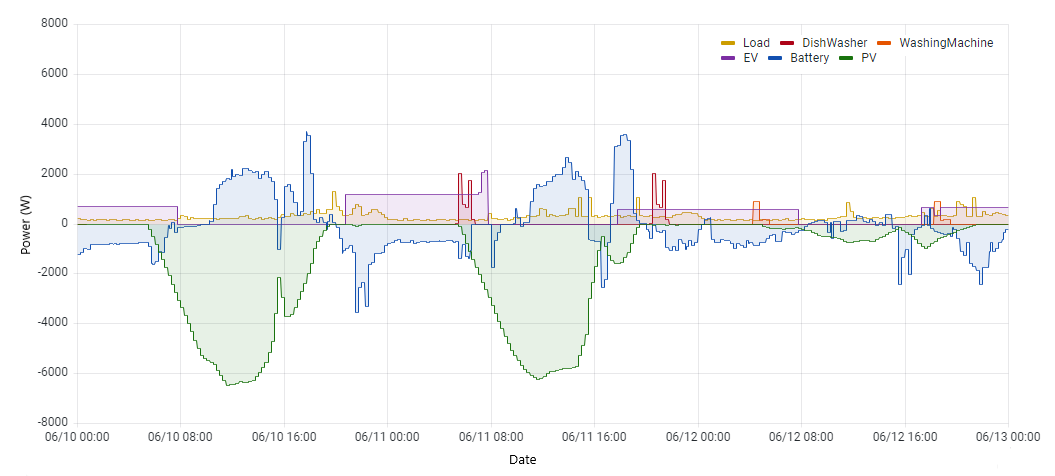} 
        \label{fig-street-10-houses-4.d}}
    \caption{Power profiles results of devices like: dish washer, washing machine, battery, electric vehicles, and photovoltaic panels from one house of the street of 10 houses in Helsinki generated before and after being optimized at \textbf{DEMKit} \cite{Hoogsteen2019DEMKit} for 3 days in June of 2017}
    \label{fig-street-10-houses-4}
\end{figure}

Figure \ref{fig-street-10-houses-4} explores the behavior of specific equipment within a single house in a 10-house street in Helsinki. The data was generated for a period of 3 days in June 2017. In Fig. \ref{fig-street-10-houses-4.a}, lighter-colored lines illustrate the power profiles of devices in their natural, uncontrolled state. Conversely, Fig. \ref{fig-street-10-houses-4.b} presents the same devices, with slightly darker lines signifying the power profiles after implementing control strategies set at 100\% control specifications. Figures \ref{fig-street-10-houses-4.c} and \ref{fig-street-10-houses-4.d} continue the analysis of these equipment profiles. The energy generated by the photovoltaic panels remains consistent across various steering approaches, as they are unrelated to the steering mechanisms. %It's worth noting that even on what are typically the sunniest days in Finland like June, such as June 12, the weather conditions were likely quite cloudy. 
The fluctuations in equipment activation and deactivation times, as depicted in Fig. \ref{fig-street-10-houses-3.a}, are also evident here in all graphs (\ref{fig-street-10-houses-4.b}, \ref{fig-street-10-houses-4.c} and \ref{fig-street-10-houses-4.d}  compared to themselves and also to the  base case in \ref{fig-street-10-houses-4.a}). The alterations to the charging pattern curve for electric vehicles with elongated charging period repeats here at \ref{fig-street-10-houses-4.b} and \ref{fig-street-10-houses-4.d}, as previously shown in Fig. \ref{fig-street-10-houses-3.b}. In Fig. \ref{fig-street-10-houses-4.c}, an interesting pattern unfolds in the battery's charging behavior. During this period, the battery remains charged with occasional new charges and does not discharge, indicating that the majority of the energy generated by the PV panels was directed to the grid, while the consumed energy was also sourced from the grid. Prices for import and export were the same and the local production was neglected by the battery control.

%%%%%%%%%%%%%%%%%%%%%%%%%%%%%%%%%%%%%%%%%%%%%%%%%%%%%%%%%%%%%%%%%%%%%%%%%%%%%%%%%%%%%%%%%%%
\section{Conclusions}\label{sec:Conclusions}
In this paper, we have examined the use of non-pricing mechanisms, such as the Profile Steering method, for performing DSM. We performed several simulations using DEMKit for a street in Helsinki, Finland, and demonstrated the influence of the Profile Steering method on peak loads, losses, and device profiles. We showed that peak-load control strategies contribute to reducing peak power and improving power flow stability, while strategies primarily based on prices may result in higher peaks and increased grid losses. 
%This study emphasizes the benefits of using the flexibility of some loads, like electric vehicles and appliances with adjustable schedules, to achieve higher grid stability. It also highlights the importance of using control strategies to manage the grid energy more efficiently when the renewable energy sources are more productive, as residential solar panels, installed in some houses, generate more power on warmer days.

%This research provides insights into the potential benefits of demand-side management in residential areas with flexible loads and PV energy generation, highlighting the role of smart control systems in optimizing energy consumption and grid performance. The findings contribute to the understanding of how demand-side management mechanisms can enhance the resilience and efficiency of local energy grids in Finland, particularly in the context of evolving energy consumption patterns and renewable energy integration.

%, this paper utilizes DEMKit, a software tool developed at the University of Twente, to illustrate the application of demand-side management mechanisms in a residential setting. DEMKit offers a model predictive control system for managing multi-energy systems. The simulation focused on a scenario involving a street with 10 houses in Helsinki, Finland, considering different occupancy and employment scenarios, along with a mix of devices such as electric vehicles, heat pumps, batteries, and photovoltaic panels.

%%%%%%%%%%%%%%%%%%%%%%%%%%%%%%%%%%%%%%%%%%%%%%%%%%%%%%%%%%%%%%%%%%%%%%%%%%%%%%%%%%%%%%%%%%%
\section*{Acknowledgment}

The authors thanks Dr. Ir. Gerwin Hoogsteen from University of Twente for helping us with using the DEMKit software. %available for us and promptly responding all of our questions.

\bibliographystyle{./IEEEtran}
\bibliography{References}

\end{document}